# Poster Abstract: Mobile Phone Based Portable Field Sensor System for Real-Time In-situ River Water Quality Monitoring During Endangered Dolphin Monitoring Surveys

Sanaullah Manzoor[1], Farhan Ahmad, Suleman Mazhar
Faculty of Computer Science
Information Technology University, Pakistan.

{sanaullah.manzoor, farhan.ahmad1, suleman.mazhar}@itu.edu.pk

## ABSTRACT

Mobile phone based potable water quality assessment device is developed to analyze and study water pollution level at Indus river. Indus river is habitat of endangered Indus river dolphin and water pollution is one of major causes of survivability threats for this specie. We tested device performance at the six locations of Lahore canal. pH of canal water deviates from the normal range of the irrigation water. In future, we will study correlation between water pollution level and habitat usage of Indus river dolphin using water quality assessment device and hydrophone array based passive acoustic monitoring (PAM) system.

## 1 INTRODUCTION

Indus river dolphin (*Platanista gangetica minor*) is declared an endangered specie under the International Union for Conservation of Nature- (IUCN) red-list due to population decline and conservation threats. Habitat has great impact on the behavior, foraging activities and social life of under-water mammals. Habitat of Indus river dolphin is Indus river which one of the longest river in Asia, spanning about 3,610 km, originates from China and merge into Arabian Sea near Thatta, Sindh province of Pakistan. Water pollution, habitat degradation, climate changes, construction of dams and barrages, monsoon flood and inadequate surface water level are the main causes of population decline. Water pollution has direct impact on the life and population distribution of cetaceans. Indus river water pollution is due to chemical discharge from textile mills, tanneries, sugar mills, wood and jute mills, and also by pulp and paper factories [1].

Alarmingly growing pollution level of Indus river water has raised questions over the habitat usage and survivability status of Indus river dolphin. To conserve endangered species like Indus river dolphin, there is a need to assess surface water impurities of habitat and its impact on population trends. There exists a strong correlation between habitat usage of a specie and water pollution level [2]. To observe this correlation, there are several water quality assessment parameters that have been suggested for the irrigation water quality monitoring, such as water pH level, conductivity, temperature, dissolved oxygen and total suspended solids [3].

In this paper, we developed a mobile-phone based device for real time in-situ river water quality monitoring in parallel with dolphin passive acoustic monitoring (PAM) survey to understand the relationship between endangered dolphin population distribution and river water quality. Water pH and temperature are selected as study parameters. Water quality monitoring device is microcontroller based system having pH and temperature sensors while PAM system consists an array of two hydrophones along with NI DAQ.

## 2 SYSTEM DESCRIPTION

To study the correlation between water pollution level and population trends of Indus river dolphin, Figure 1 shows an overview of the system. System consists of two major components, firstly water quality assessment device to monitor water impurity level and secondly hydrophone array to study passive acoustic monitoring. Water quality monitoring system consists of a sensor node which is a microcontroller board provides an interfacing of pH and temperature sensors with Andriod application using Bluetooth module. The reason behind mobile phone inclusion in the design is that it will not only reduces addition cost of GPS (Global Positioning System) module but also enables the device to upload data on the database server to process and analyze in the real-time.

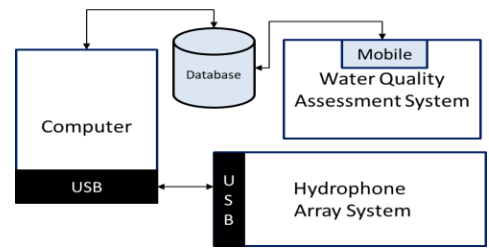

**Figure 1: Portrays system overview for the study of water pollution impact on habitat of endangered Indus river dolphin.**

## 2.1 Water Quality Assessment Device

Block diagram of water quality assessment device is shown in Figure 2. Device has Arduino UNO along with Bluetooth HC-05 module and two water quality measuring sensors namely as pH and temperature sensors. Device gets pH and temperature values and transfer it to mobile phone through Bluetooth interface. Mobile phone stores sensor data along with the system date and location to the database. Figure 3 shows components of the system, a) shows temperature and pH sensor, b) is circuitry of water quality testing device, c) screenshot of Android application and d) shows PAM system, two hydrophones array with NI DAQ.

pH value indicates alkaline or acidity of water or fluid. Range of pH varies from 0 to 14. Range 0-6 are acidic representation while 8-14 value pH indicates alkaline characteristics of the water. Normal range of irrigation water is 6.5 to 8.4 [4]. Accuracy error of pH sensor (E-201-C) is $\pm 0.1$ pH ($0°C$) and measuring rage is 0-14 pH. Literature shows that irrigation water temperature varies form 17° C-19° C in the winter while 27° C-29° C in the summer season [4]. Temperature study is important because there is inverse relationship between temperature and dissolved oxygen (DO), increase in temperature level decreases dissolved oxygen contents in the water. Temperature sensor (K-201) has measuring range from 0-60℃. Android mobile application is developed to get data from sensors at remote locations effectively and store it into database for further use. Database has these following attributes: date, time, longitude, latitude, pH value and water temperature value.

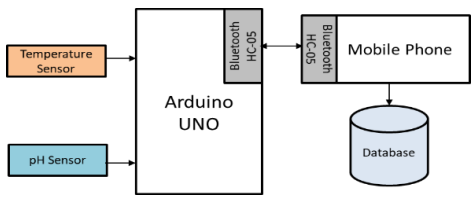

**Figure 2: Shows water quality monitoring system. Arduino board is responsible for interfacing sensors with mobile phone.**

## 2.2 Hydrophone array based PAM system

PAM system is used to localize and track underwater cetaceans. Figure 3 shows PAM system. Two hydrophones attached with NI DAQ (NI USB PXI Express 6366) for high rate data sampling.

## 3 RESULTS AND DISCUSSION

We tested water quality monitoring device at Lahore canal. Experiments are performed at 6 samples different locations of canal and at each point after 3 minutes of arrival, pH, temperature, longitude and latitudes are measured. Figure 4 shows experimental results. pH value varies from 5.33 to 6.38, while temperature from 25.9° to 28.3°C. pH readings indicates acidic characteristics of canal water and it is below from normal range (6.5 to 8.4) of irrigation water, while temperature value is in normal range (27° C-29° C).

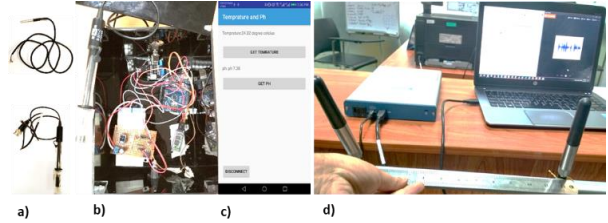

**Figure 3: Presents system components: a) Temperature and pH sensor, b) Water quality assessment device, c) screenshot of Android application. And d) An array of two hydrophones with NI DAQ.**

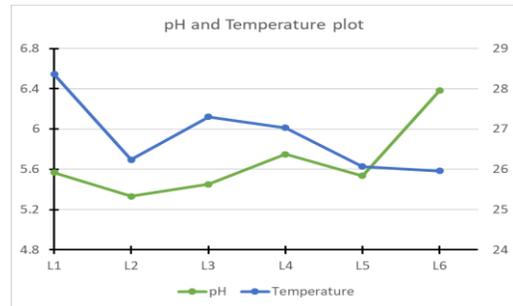

**Figure 4: Plot of pH and temperature from six locations (L1-L6) of Lahore canal.**

## 4 CONCLUSION AND FUTURE WORK

We presented a portable mobile phone-based river water quality testing device to measure the water pollution level at habitat of endangered Indus river dolphin. Device is tested at six locations of Lahore canal and is capable of monitoring pH and temperature characteristics of water. Only pH of canal water deviates from the normal range of the irrigation water. Mobile phone based design helped us to collect data from remote locations effectively through uploading real-time sensor data to database server. In future, we will study the correlation between Indus river dolphin population trends and water pollution level at Indus river using two hydrophones based PAM system along with water quality testing device.